\documentclass{aa}  

\usepackage{graphicx}
\usepackage{txfonts}
\usepackage{hyperref}
\usepackage{siunitx}
\usepackage{color}

\usepackage{amsmath}

\def\reff{R_{\mathrm{e}}}
\def\lowre{r_{\mathrm{e}}}
\def\mh{m_{\mathrm{h}}}
\def\sap{s_{\mathrm{ap}}}
\def\sapobs{s_{\mathrm{ap}}^{(\mathrm{obs})}}

\def\mstar{M_*}
\def\msps{M_*^{(\mathrm{sps})}}
\def\sstar{\Sigma_*^{(\mathrm{sps})}}
\def\mstarobs{M_*^{(\mathrm{obs})}}

\def\asps{\alpha_{\mathrm{sps}}}
\def\abarsps{\bar{\alpha}_{\mathrm{sps}}}
\def\eps{\epsilon}

\def\zsource{z_{\mathrm{s}}}
\def\zlens{z_{\mathrm{g}}}

\def\zsourcei{z_{\mathrm{s},i}}

\def\gammadm{\gamma_{\mathrm{DM}}}

\def\mhalo{M_{\mathrm{h}}}
\def\mtwoh{M_{200}}

\def\gammapl{\gamma_{\mathrm{PL}}}
\def\gammaplobs{\gamma_{\mathrm{PL}}^{(\mathrm{obs})}}
\def\nlens{n_{\mathrm{lens}}}
\def\nfgd{n_{\mathrm{g}}}
\def\nbkg{n_{\mathrm{s}}}

\def\tein{\theta_{\mathrm{E}}}
\def\teinest{\theta_{\mathrm{E}}^{(\mathrm{est})}}

\def\teinobs{\theta_{\mathrm{E}}^{(\mathrm{obs})}}

\def\hyperpars{\boldsymbol{\eta}}

\def\psilens{\boldsymbol{\psi}_\mathrm{g}}
\def\psilensein{\boldsymbol{\psi}_\mathrm{g}^{(\mathrm{E})}}
\def\psilensi{\boldsymbol{\psi}_{\mathrm{g},i}}

\def\prlens{{\rm P}_\mathrm{g}}

\def\prsourceeff{{\rm P}_\mathrm{s}^{(\mathrm{eff})}}

\def\prsleff{{\rm P}_\mathrm{{SL}}^{(\mathrm{eff})}}

\def\psel{{\rm P}_\mathrm{sel}}
\def\pfind{{\rm P}_\mathrm{find}}

\def\sigmaap{\sigma_\mathrm{{ap}}}
\def\sigmaapobs{\sigma_\mathrm{{ap}}^{(\mathrm{obs})}}

\def\data{\mathbf{d}}
\def\datai{\mathbf{d}_i}

\def\Sref#1{Section~\ref{#1}\xspace}
\def\Fref#1{Figure~\ref{#1}\xspace}
\def\Tref#1{Table~\ref{#1}\xspace}
\def\Eref#1{Equation~\ref{#1}\xspace}

\def\pr{{\rm P}}

\defcitealias{Son24}{Paper~I}
\setcitestyle{notesep={}}

\begin{document}

   \title{The SLACS strong lens sample debiased. II. Lensing-only constraints on the stellar IMF and dark matter contraction in early-type galaxies}
   \titlerunning{SLACS debiased, II}
   \authorrunning{Sonnenfeld}

   \author{Alessandro Sonnenfeld\inst{\ref{sjtu1},\ref{sjtu2},\ref{sjtu3}}
          }

   \institute{
Department of Astronomy, School of Physics and Astronomy, Shanghai Jiao Tong University, Shanghai 200240, China\\
              \email{sonnenfeld@sjtu.edu.cn}\label{sjtu1} \and
Shanghai Key Laboratory for Particle Physics and Cosmology, Shanghai Jiao Tong University, Shanghai 200240, China\label{sjtu2} \and
Key Laboratory for Particle Physics, Astrophysics and Cosmology, Ministry of Education, Shanghai Jiao Tong University, Shanghai 200240, China\label{sjtu3}
             }

   \date{}

 
  \abstract
    {
The Sloan Lens ACS (SLACS) is the best studied sample of strong lenses to date.
Much of our knowledge of the SLACS lenses has been obtained by combining strong lensing with stellar kinematics constraints. 
However, interpreting stellar kinematics data is difficult: it requires reconstructing the three-dimensional structure of a galaxy and the orbits of its stars.
For SLACS, the problem is exacerbated by its selection function, which caused lenses with a larger observed velocity dispersion to be overrepresented.
In this work we pursued an alternative approach to the study of galaxy structure with SLACS, based purely on gravitational lensing data. 
The primary goal of this study is to constrain the stellar population synthesis mismatch parameter $\asps$, quantifying the ratio between the true stellar mass of a galaxy and that obtained with a reference stellar population synthesis model, and the efficiency of the dark matter response to the infall of baryons, $\eps$.
We combined Einstein radius measurements from the SLACS lenses with weak lensing information from their parent sample, while accounting for selection effects.
The data can be fit comparatively well by a model with $\log{\asps}=0.22$ and $\eps=0$, corresponding to an IMF slightly lighter than Salpeter and no dark matter contraction, or $\log{\asps}=0$ and $\eps=0.8$, equivalent to a Chabrier IMF and almost maximal contraction.
Selection effects, if not modelled, produce a shift in the joint posterior probability that is larger than the uncertainty.
The degeneracy between $\asps$ and $\eps$ could be broken with lensing-only measurements of the projected density slope, but existing data are completely inconsistent with our model. We suspect systematic errors in the measurements to be at the origin of this discrepancy.
Number density constraints would also help break the degeneracy.
Because of selection effects, SLACS lenses have a larger velocity dispersion than galaxies with the same projected mass distribution, and their velocity dispersion is overestimated. These two biases combined produce a $5\%$ upward shift in the observed velocity dispersion.
}

   \maketitle
%

\section{Introduction}\label{sect:intro}

Strong gravitational lensing is one of the most reliable methods for studying the mass distribution of galaxies: it can provide a measurement of the projected mass of a lens galaxy with a precision and accuracy of a few percent, regardless of the dynamical state of the lens.
The best studied sample of lenses so far is the Sloan Lens ACS (SLACS) survey \citep{Bol++06}, consisting of about a hundred strong lenses, most of which are massive early-type galaxies.
The standard strategy for studying the inner structure of the SLACS lenses has so far been to combine information from strong lensing and stellar kinematics. 
Joint lensing and stellar dynamics analyses have revealed that SLACS lenses have, on average, a total density profile slightly steeper than isothermal \citep{Koo++06,Aug++10a,Bar++11}, and that the steepness of the density profile correlates positively with the stellar mass density \citep{Son++13b}.
Lensing and dynamics have also been used to determine the relative contribution of stars and dark matter to the total mass budget of massive galaxies \citep{Tre++10,Aug++10b,Son++15,Pos++15,Sha++21,She++25}.
According to a large part of these investigations, the stars appear to have a relatively large mass-to-light ratio, similar to that predicted by a Salpeter stellar initial mass function (IMF), while the dark matter mass enclosed within the inner regions seems to be smaller than that predicted by numerical simulations \citep[see e.g. Fig. 11 of][]{Sha++21}.

Interpreting the stellar kinematics data of a galaxy is difficult.
It requires modelling its three-dimensional mass structure and the orbits of its stars, which are two properties that are not directly observable. 
While some of the degeneracies affecting dynamical models can be broken with the help of integral field spectroscopic data, for most of the SLACS lenses the only stellar kinematics information currently available comes from Sloan Digital Sky Survey (SDSS) single-aperture spectra.
This scarcity of spatially resolved spectroscopic data has forced past investigations to rely on simplifying assumptions: nearly all of the studies mentioned above have assumed spherical symmetry, and some of them have asserted orbital isotropy \citep{Aug++10a,Son++15}.
One notable exception is the work of \citet{Bar++11}, who fitted models with axial symmetry and orbits described by a two-integral distribution function to spatially resolved kinematics data of 16 lenses.
\citet{Bar++11} modelled the radial density profile of the lenses as a power-law, $\rho(r)\propto r^{-\gamma}$ and measured the density slope $\gamma$ for each of the lenses. 
Their values of $\gamma$ are systematically lower, by approximately $0.05$, than those obtained by \citet{Aug++10a} with an isotropic spherical Jeans model. 
Such a discrepancy is larger than the current uncertainty on the average $\gamma$ of the SLACS lens sample, and should therefore warn against assuming spherical symmetry or isotropy when studying the SLACS lenses.

The problem of analysing SLACS stellar kinematics data is made worse by selection effects. 
SLACS lenses were selected on the basis of the observed velocity dispersion \citep{Bol++06}. 
This caused lenses with a larger observed velocity dispersion to be overrepresented, with respect to a purely mass-selected sample \citep{Son24}.
In order for a dynamical analysis to accurately capture the mass and orbital structure of a SLACS lens, these selection effects must be taken into account. 
So far, this has only been done by \citet[][, hereafter Paper I]{Son24}, albeit in the simplified context of spherical and isotropic models.
Since the velocity dispersion correlates with the three-dimensional structure and orbital anisotropy of a galaxy, a more complex and realistic dynamical model of a SLACS lens would require a non-trivial prior on these quantities: for instance, models with a larger radial anisotropy and elongation along the line of sight would have to be upweighted.

In this paper we re-analysed the SLACS sample with a more conservative approach: we focused exclusively on projected quantities, which are directly observable, and used only gravitational lensing data as constraints on the mass distribution.
We combined strong lensing measurements of SLACS with weak lensing observations obtained for the parent galaxy sample from which the lenses were drawn, while explicitly accounting for selection effects.
The primary goal of this work is to put constraints on the stellar mass-to-light ratio of massive galaxies and on their inner dark matter density profile.
We want to determine to what extent can gravitational lensing data, on its own, support previous claims of a non-universal IMF in early-type galaxies, and whether it provides evidence for the contraction of dark matter, a phenomenon ubiquitously observed in hydrodynamical simulations at the mass scales of the SLACS lenses \citep{Gne++04,Duf++10,Sch++15,Cau++20}, but for which there is still little experimental evidence.

Our secondary goal is to better understand how the velocity dispersion of the SLACS lenses differs from that of massive galaxies of the same mass, and to corroborate the claim made in Paper I that the values of the observed velocity dispersion are overestimated.
This is an important point because the SLACS lenses, with their stellar kinematics data, are being used in support of time delay measurements of the Hubble constant \citep{Bir++20}, as a source of information on the internal structure of strong lenses. If the SLACS lenses are biased in a unique way with respect to the general population of galaxies, they might also be biased with respect to the time-delay lenses, and therefore bias the measurement of the Hubble constant.

This paper is structured as follows.
In \Sref{sect:methods} we describe the methods and data used for the analysis.
In \Sref{sect:results} we show the results. We discuss and conclude in \Sref{sect:discuss}.
Throughout this work we assumed a flat $\Lambda$ cold dark matter model, with $H_0 = 70\,\mathrm{km}\,\mathrm{s}^{-1}\,\mathrm{Mpc}^{-1}$, and $\Omega_m = 0.3$.
All masses are in Solar units, lengths are in units of kpc, and velocities are in ${\rm km}$~s$^{-1}$.

\section{Methods}\label{sect:methods}

\subsection{Analysis formalism}\label{ssec:formalism}

Our analysis method follows in large part that of Paper I. The starting point is a simplified version of the fundamental equation of statistical strong lensing,
\begin{equation}\label{eq:one}
\prsleff(\psilens,\zsource) \propto \prlens(\psilens|\hyperpars)\prsourceeff(\zsource|\hyperpars)\psel(\psilens,\zsource|\hyperpars),
\end{equation}
which describes the probability distribution of lens-source pairs (left-hand side) as the product of the distribution of foreground galaxies $\prlens$, the empirically-derived effective distribution of background sources $\prsourceeff$, and the lens selection probability $\psel$. In the equation above, $\psilens$ indicates the set of parameters describing a foreground galaxy, $\zsource$ is the source redshift, while $\hyperpars$ is the ensemble of population-level parameters, also referred to as hyper-parameters of the model.
\Eref{eq:one} was derived in Paper I by means of a key approximation: the probability of a galaxy-source pair to give rise to a detectable strong lens separates cleanly into a factor that depends purely on geometry and another factor that depends purely on the source brightness. That approximation allowed us to eliminate any explicit dependence on the source brightness from \Eref{eq:one}.

The goal of the analysis is to infer the population-level parameters $\hyperpars$ given the data $\data$ of the SLACS lenses. The posterior probability of $\hyperpars$ is
\begin{equation}
\pr(\hyperpars|\data) \propto \pr(\hyperpars)\pr(\data|\hyperpars),
\end{equation}
and the likelihood is the product over all of the SLACS lenses:
\begin{equation}
\pr(\data|\hyperpars) = \prod_i \pr(\datai|\hyperpars).
\end{equation}
To evaluate each factor, it is necessary to marginalise over all possible values of the individual lens parameters and source redshift, the prior of which is given by the distribution of \Eref{eq:one}:
\begin{equation}\label{eq:integral}
\pr(\datai|\hyperpars) = \int d\psilensi d\zsourcei \pr(\datai|\psilensi,\zsourcei,\hyperpars)\prsleff(\psilensi,\zsourcei|\hyperpars).
\end{equation}

\subsection{Data}

The sample and data used are the same as in Paper I. These are $59$ early-type galaxy strong lenses from the SLACS sample, each with the following measurements:
\begin{itemize}
\item The lens and source spectroscopic redshift, $\zlens$ and $\zsource$.
\item The stellar population synthesis-based stellar mass, $\mstarobs$, based on a Chabrier IMF and a de Vaucouleurs stellar profile, and its uncertainty.
\item The half-light radius $\reff$.
\item The Einstein radius, $\teinobs$, with a conservative $5\%$ uncertainty. 
\item The line-of-sight stellar velocity dispersion from SDSS, $\sigmaapobs$, and its uncertainty.
\end{itemize}
These data were taken from \citet{Aug++09}. Uncertainties on redshifts and half-light radii are small and were ignored in the analysis.


\subsection{Mass model}

We adopted a two-component mass model, consisting of a stellar bulge and a dark matter halo.
The stars follow a de Vaucouleurs profile, described by two parameters: the total stellar mass, $\mstar$, and the half-light radius, $\reff$.
Additionally, we considered the stellar population synthesis stellar mass, $\msps$, which we defined as the stellar mass that an observer would infer by fitting the stellar population synthesis model of \citet{Aug++09} to perfect photometric data. 
The true stellar mass $\mstar$ is related to $\msps$ by means of the stellar population synthesis mismatch parameter, $\asps$, defined as
\begin{equation}
\asps = \frac{\mstar}{\msps}.
\end{equation}
This means that, if the \citet{Aug++09} stellar population synthesis model is accurate, then $\asps=1$. The stellar population mismatch parameter is one of the main quantities that we seek to constrain in this work.

To describe the dark matter density profile, we emulated the effect of the gravitational contraction resulting from the infall of baryons.
The idea is that the dark matter and the baryonic matter that eventually formed stars followed initially the same distribution. Subsequently, the baryons sank to the centre due to dissipative processes, altering the gravitational potential, and the dark matter particles responded by shrinking their orbits.
We modelled this process under the approximation of adiabatic contraction of particles in circular orbits, as originally suggested by \citet{Blu86}.
Although neither the contraction of dark matter is truly adiabatic nor the orbits are circular, this is a widely used approach in semi-analytic models of dark matter halos, and produces results that are qualitatively similar to those of hydrodynamical simulations \citep[see e.g.][]{Cau++20}.

For a collisionless dark matter particle in a circular orbit at radius $r$, if the total mass $M(r)$ enclosed within the sphere with radius $r$ varies slowly with respect to the orbital time, then the quantity $rM(r)$ is conserved. If $r_i$ and $r_f$ are respectively the initial (pre-contraction) and final radius of a dark matter particle, then the conservation of $rM(r)$ implies that
\begin{equation}\label{eq:adcontr0}
r_iM_{\mathrm{DM},i}(r_i)\left(1 + \frac{M_*}{\mhalo}\right) = r_f\left[M_{\mathrm{DM},f}(r_f) + M_*(r_f)\right].
\end{equation}
In the above equation, $\mhalo$ is the total halo mass, while $M_{\mathrm{DM},i}(r)$ and $M_{\mathrm{DM},f}$ indicate the initial and final dark matter density profiles. Shells at different $r$ do not cross each other, so $M_{\mathrm{DM},i}(r_i) = M_{\mathrm{DM},f}(r_f)$, and \Eref{eq:adcontr0} becomes
\begin{equation}\label{eq:adcontr1}
r_iM_{\mathrm{DM},i}(r_i)\left(1 + \frac{M_*}{\mhalo}\right) = r_f\left[M_{\mathrm{DM},i}(r_i) + M_*(r_f)\right].
\end{equation}
This equation describes a dark matter halo that responds solely to the baryons that end up forming the central galaxy.
In reality, star formation is not $100\%$ efficient, and part of the baryons involved in the star formation process are ejected by winds, supernovae or active galactic nuclei. Any gas that is ejected from the inner regions would cause the dark matter halo to expand, rather than contract. Consequently, we expect the model described above to be an upper limit to the amount of contraction. In order to allow for a range of contraction scenarios, we modified \Eref{eq:adcontr1} by introducing an adiabatic contraction efficiency parameter, $\epsilon$:
\begin{align}\label{eq:adcontr2}
r_iM_{\mathrm{DM},i}(r_i) & \left(1 + \frac{M_*}{\mhalo}\right) = \nonumber \\
& r_f\left[M_{\mathrm{DM},i}(r_i)\left(1 + (1-\epsilon)\frac{M_*}{\mhalo}\right) + \epsilon M_*(r_f)\right].
\end{align} 
In words, \Eref{eq:adcontr2} is equivalent to a model in which dark matter responds to the contraction of only a fraction $\epsilon$ of the stellar mass of the galaxy.
This parameterisation of the contraction efficiency is an alternative to the widely-adopted one of \citet{Dut++07}, which is based on first obtaining the contraction factor $r_f/r_i$ of each shell from \Eref{eq:adcontr1}, and then rescaling it by taking its power $(r_f/r_i)^\nu$.
Our model is easier to implement numerically and, we believe, more intuitive.

Given the initial dark matter distribution and the final stellar mass distribution, \Eref{eq:adcontr0} allows us to determine the final dark matter profile, by solving for $r_i$ at any value of $r_f$ (or vice-versa).
We allowed the value of $\epsilon$ to vary continuously from $\epsilon=1$, which corresponds to the maximally contracted case, to $\epsilon=0$, for which the solution to \Eref{eq:adcontr2} is simply $r_f=r_i$.
We modelled the initial dark matter distribution with a Navarro, Frenk \& White (NFW) profile. We set $\mtwoh=\mhalo$, where $\mtwoh$ is the mass within a sphere with average density equal to $200$ times the critical density of the Universe, and adopted the following mass-concentration relation from \citet{D+M14} to determine the initial scale-radius:
\begin{equation}\label{eq:dutmac}
\log{c_{200}} = 0.905 - 0.101(\log{\mtwoh} - 12 + \log{h}),
\end{equation}
with $h=0.7$.
We neglected scatter in concentration at fixed halo mass, to keep the dimensionality of the problem low.

\begin{figure*}
\sidecaption
 \includegraphics[width=12cm]{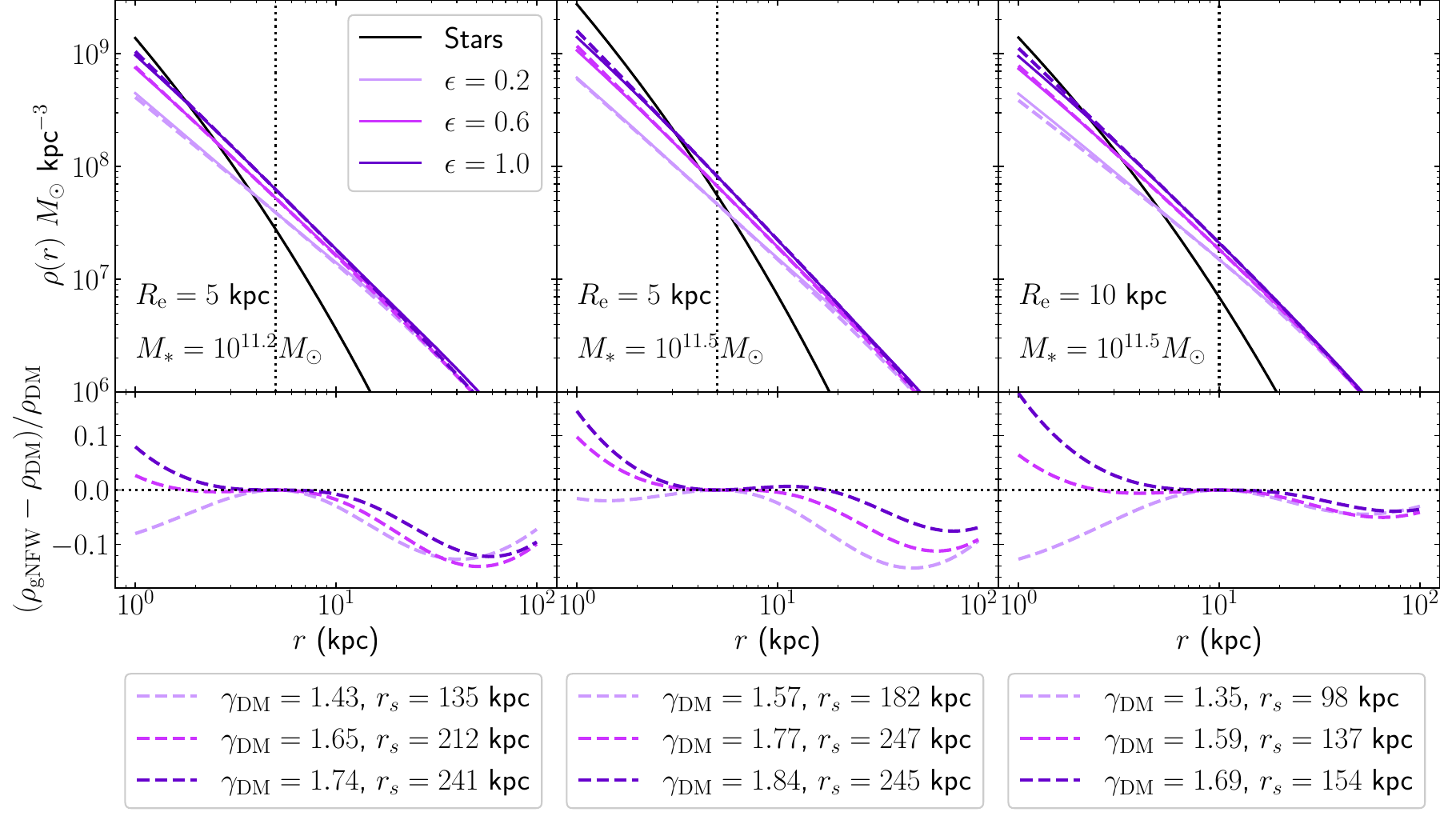}
  \caption{Stellar and dark matter density profile of example model galaxies.
In each top panel, the solid black line shows the stellar mass density, which follows a spherically de-projected de Vaucouleurs profile, with half-light radius indicated by the vertical dotted line. Solid coloured lines are dark matter profiles obtained by applying the adiabatic contraction prescription of \Eref{eq:adcontr2}, with different values of the contraction efficiency parameter $\epsilon$. Dashed lines are gNFW profile approximation of the contracted profile.
The bottom panels show the relative difference between the gNFW and the corresponding adiabatically contracted profile. The legends at the bottom indicate the values of $\gammadm$ and $r_s$ of each gNFW model.
All galaxies have a halo mass of $\mtwoh = 10^{13}M_\odot$. Stellar mass and half-light radius vary as indicated on the top panels.
}
\label{fig:adcontr}
\end{figure*}

The dark matter profile that is obtained by solving \Eref{eq:adcontr2} is a non-analytic function, which makes it cumbersome to use to predict lensing-related quantities. In order to speed up computations, we approximated the resulting profile with a generalised Navarro, Frenk \& White (gNFW) model:
\begin{equation}
\rho(r) \propto \dfrac{1}{r^{\gammadm}\left(1 + \frac{r}{r_s}\right)^{3-\gammadm}}.
\end{equation}
A gNFW profile has three degrees of freedom: the mass normalisation, the scale radius $r_s$ and the inner density slope $\gammadm$.
To determine these quantities, we fixed the virial mass to be the same as that of the uncontracted halo, and set $r_s$ and $\gammadm$ so as to match the density and the logarithmic density slope, $d\log{\rho}/d\log{r}$, at $r=\reff$.
\Fref{fig:adcontr} shows examples of dark matter density profiles determined by solving \Eref{eq:adcontr2}, along with their gNFW approximations, for a few values of $\epsilon$, $M_*$, and $\reff$, at fixed halo mass.
For a given $\epsilon$, dark matter profiles get steeper with increasing stellar mass and with decreasing half-light radius. The gNFW approximation matches the original dark matter profile to better than $10\%$, at most radii.

In summary, we described the dark matter halo by means of two parameters: $\mtwoh$ and $\epsilon$. To compute its lensing properties, we used the gNFW profile corresponding to the values of $\mtwoh$, $\epsilon$, $\mstar$ and $\reff$ of the galaxy. 
For positive values of $\epsilon$, the central dark matter density correlates positively with the stellar mass and negatively with the half-light radius.


\subsection{Population distribution}\label{ssec:popmodel}

Our inference method requires us to model the distribution of foreground galaxies and background sources that generate the SLACS lenses: $\prlens$ and $\prsourceeff$ in \Eref{eq:one}. The former describes the parent population of galaxies among which SLACS lenses were searched for, which are early-type galaxies in the SDSS spectroscopic sample. The foreground galaxy parameters that are relevant for the problem are 
\begin{equation}
\psilens \equiv \{\zlens,m_*,\lowre,\asps,\mh,\epsilon,\sap\},
\end{equation}
where we introduced the following compact notation:
\begin{align}
m_* & \equiv \log{\frac{\msps}{M_\odot}}, \\
\lowre & \equiv \log{\frac{\reff}{\mathrm{kpc}}}, \\
\mh & \equiv \log{\frac{\mtwoh}{M_\odot}}, \\
\sap & \equiv \log{\frac{\sigmaap}{\mathrm{km}\,\mathrm{s}^{-1}}}.
\end{align}
We chose the following form for their distribution:
\begin{align}\label{eq:popdist}
\prlens(\psilens) = & \mathcal{M}(z,m_*)\mathcal{R}(\lowre|m_*)\delta(\asps - \abarsps) \mathcal{H}(\mh|m_*) \times \nonumber \\
& \delta(\epsilon - \bar{\epsilon})\mathcal{S}(\sap|m_*,\lowre,\mh).
\end{align}

We set the factors $\mathcal{M}$ and $\mathcal{R}$, describing the distribution in stellar mass and half-light radius, to be the same as in Paper I. The former is a Schechter function with a redshift-dependent truncation, while the latter is a Gaussian in $\lowre$ with the mean that scales quadratically with $m_*$.
We asserted $\asps$ to be the same across all population.
For the distribution in halo mass, we adopted the same model used by \citet{Son++18}:
\begin{equation}\label{eq:halodist}
\mathcal{N}(\mh|m_*) = \mathcal{N}_{\mh}(\mu_\mathrm{h}(m_*),\sigma_{\mathrm{h}}^2),
\end{equation}
where the notation $\mathcal{N}_x(\mu,\sigma^2)$ indicates a Gaussian distribution in $x$ with mean $\mu$ and variance $\sigma^2$, and
\begin{equation}\label{eq:muhalo}
\mu_{\mathrm{h}} = \mu_{\mathrm{h,0}} + \beta_{\mathrm{h}}(m_* - 11.3).
\end{equation}
\citet{Son++18} measured the distribution in halo mass of the parent population of the SLACS lenses, by fitting the model of \Eref{eq:halodist} to weak lensing observations from the Hyper Suprime-Cam Subaru Strategic Program \citep[HSC SSP][]{Aih++18a,Aih++18b}.
They constrained the parameters of \Eref{eq:halodist} and \Eref{eq:muhalo} to be $\mu_{\mathrm{h,0}} = 13.04\pm0.04$, $\beta_{\mathrm{h}} = 1.48\pm0.15$ and $\sigma_{\mathrm{h}} = 0.31\pm0.04$, with very small covariance between them. 
We used these measurements as a prior on the three parameters, thus incorporating weak lensing information on the halo mass distribution of the foreground galaxies.
We approximated this prior as a trivariate Gaussian with diagonal covariance matrix.
One implicit assumption is that the halo mass distribution is independent of the contraction parameter $\epsilon$. This is justified by the fact that \citet{Son++18} fitted both adiabatically contracted and NFW models to the weak lensing data, finding negligible differences in the inferred halo mass distribution parameters between the two models.
As with the parameter $\asps$, we also asserted that $\epsilon$ is the same among all galaxies.
For simplicity of notation, we replace $(\abarsps,\bar{\epsilon})$ with $(\asps,\epsilon)$ from here on.

The last factor in \Eref{eq:popdist} is a term describing the distribution in velocity dispersion of parent population galaxies. This term is needed because the SLACS selection function depends explicitly on the observed velocity dispersion.
In Paper I we predicted $\sigmaap$ by means of dynamical models, assuming a spherical mass distribution and isotropic orbits.
Here we take a less assumption-heavy approach, by simply proposing an empirical scaling relation between velocity dispersion and stellar mass, size and halo mass:
\begin{equation}\label{eq:sigmadist}
\mathcal{S}(\sap) = \mathcal{N}_{\sap}(\mu_\sigma(m_*,\lowre,\mh),\sigma_\sigma^2),
\end{equation}
with
\begin{align}\label{eq:sigmamu}
\mu_\sigma(m_*,\lowre,\mh) = & \mu_{\sigma,0} + \beta_\sigma(m_* - 11.3) +\xi_\sigma(\lowre - \mu_R(m_*)) +\nonumber \\
 &  \nu_\sigma(\mh - \mu_{\mathrm{h}}(m_*)).
\end{align}
In the above equation, $\mu_R$ is the average $\lowre$ for galaxies of a given stellar mass, which we set to the quadratic mass-size relation of \citet{H+B09}.
\Eref{eq:sigmadist} can be thought of as an extension of the fundamental plane relation that includes the halo mass, so we refer to it as the fundamental hyper-plane.

The factor $\prsourceeff$ in \Eref{eq:one} describes an effective redshift distribution of background sources that can be lensed into detectable multiple images, averaged over the source brightness. 
We used the same distribution as in Paper I, which is a Gaussian in $\zsource$:
\begin{equation}
\prsourceeff(\zsource) = \mathcal{N}_{\zsource}(\mu_{\zsource},\sigma_{\zsource}^2).
\end{equation}

\subsection{Selection function}

The model for the selection probability term, $\psel$, is the same as in Paper I.
It is the product between the probability of detection of a lens-source pair and the probability of finding a lens in the SLACS data, given that it is detectable:
\begin{equation}
\psel(\psilens,\zsource) = g(\psilens,\zsource) \pfind(\psilens,\zsource).
\end{equation}
The factor $g$, which describes the probability of detection, is proportional to the strong lensing cross-section for a reference background source. 
Because the detection of a SLACS strong lens involved both spectroscopy and photometry, the choice of reference source requires defining both its surface brightness and spectrum.
We chose as reference a point source with intrinsic broadband flux equal to the detection limit of the photometric data used for imaging follow-up, and emission line flux equal to a third of the detection limit for spectroscopy. This is the same definition used in Paper I, where we found that the exact choice of reference source did not affect the results.
We then computed $g$ by simulating the production of photometric and spectroscopic data of lensed sources, as a function of the lens parameters.
As stated in section \ref{ssec:formalism}, and explained in greater detail in Paper I, we made the assumption that the lensing cross-section for sources different than the reference are simply obtained by rescaling $g$ by a factor that depends on the source spectral energy distribution, and that is implicitly included in the source distribution $\prsourceeff$.

The lens finding probability $\pfind$ describes a selection in the estimated Einstein radius that was applied by \citet{Bol++06} to prioritise the photometric follow-up of spectroscopically detected lens candidates.
Following Paper I, we modelled it as
\begin{equation}
\pfind(\teinest) = \frac{1}{1 + \exp{\left[-a(\teinest - \theta_0)\right]}}.
\end{equation}
where $\teinest$ is the Einstein radius of a singular isothermal sphere with velocity dispersion equal to the observed one: 
\begin{equation}\label{eq:teinest}
\teinest = 4\pi\left(\frac{\sigmaapobs}{c}\right)^2\frac{D_{\mathrm{ds}}}{D_{\mathrm{s}}}.
\end{equation}

\subsection{Practical implementation}

The hyper-parameters of the model are
\begin{equation}
\hyperpars = \{\asps,\epsilon,\mu_{\mathrm{h},0},\beta_{\mathrm{h}},\sigma_{\mathrm{h}},\mu_{\sigma,0},\beta_\sigma,\xi_\sigma,\nu_\sigma,\sigma_\sigma,\mu_{\zsource},\sigma_{\zsource},\theta_0,a\}.
\end{equation}
\Tref{tab:results} provides a brief description of each of them.
In order to obtain the posterior probability $\pr(\hyperpars|\data)$, we need to evaluate integrals of the kind of \Eref{eq:integral}. With out parameterisation, \Eref{eq:integral} becomes
\begin{align}
\pr(\data|\hyperpars) = & \int d\mh dm_* d\sap \pr(\teinobs|m_*,\lowre,\mh,\asps,\epsilon) \pr(m_*^{(\mathrm{obs})}|m_*) \times \nonumber \\
& \pr(\sap^{(\mathrm{obs})}|\sap)\prsleff(\zlens,m_*,\lowre,\mh,\sap,\zsource|\hyperpars),
\end{align}
where we integrated over the lens and source redshift, the likelihood of which is a Dirac delta function.
Then, we approximated the likelihood in $\sap$ as a Gaussian, which allowed us to compute the following integral over $\sap$,
\begin{align}
& I_\sigma(m_*,\lowre,\mh,\hyperpars) = \int d\sap \pr(\sapobs|\sap)\mathcal{S}(\sap|m_*,\lowre,\mh,\hyperpars) = \nonumber \\
& \frac{1}{\sqrt{2\pi(\sigma_\sigma^2 + \Delta \sap^2)}}\exp{\left\{-\frac{(\mu_\sigma(m_*,\lowre,\mh) - \sapobs)^2}{2(\sigma_\sigma^2 + \Delta\sap^2)}\right\}},
\end{align}
where $\Delta\sap$ is the observational uncertainty associated with $\sapobs$.
We are left with the following integral over the stellar and halo mass:
\begin{align}\label{eq:integral3}
\pr(\data|\hyperpars) = \int & d\mh dm_* \pr(m_*^{(\mathrm{obs})}|m_*) \pr(\teinobs|m_*,\lowre,\mh,\asps,\epsilon)\times \nonumber \\
& \mathcal{M}(\zlens,m_*) \mathcal{R}(\lowre|m_*) \mathcal{H}(\mh|m_*,\hyperpars) I_\sigma(m_*,\lowre,\mh,\hyperpars) \times \nonumber \\
& g(\psilensein,\zsource) \pfind(\sapobs,\zlens,\zsource). 
\end{align}
We computed these integrals via Monte Carlo integration and importance sampling, and sampled the posterior probability with a Markov Chain Monte Carlo.

In order for $\prsleff$ to represent a proper probability distribution, it needs to be normalised:
\begin{equation}\label{eq:pslnorm}
\int d\psilens d\zsource \prsleff(\psilens,\zsource|\hyperpars) = 1.
\end{equation}
For each draw of values of the hyper-parameters, we computed integrals of the kind of \Eref{eq:pslnorm} via Monte Carlo integration, and rescaled the factor $g$ in \Eref{eq:integral3} accordingly.

\begin{table*}
\caption{Results.}
\label{tab:results}
\begin{tabular}{ccccl}
Parameter & Prior & Inference & Description \\
\hline
\hline
$\log{\asps}$ & $U(0,0.3)$ & $0.07\pm0.07$ & Logarithm of the stellar population synthesis mismatch parameter \\
$\epsilon$ & $U(0,1)$ & $0.49\pm0.30$ & Dark matter contraction efficiency parameter \\
$\mu_{\mathrm{h},0}$ & $\mathcal{N}(13.04, 0.04)$ & $13.03\pm0.04$ & Mean $\log{\mtwoh}$ at $\log{\msps}=11.3$ \\
$\beta_{\mathrm{h}}$ & $\mathcal{N}(1.48, 0.15)$ & $1.55\pm0.13$ & Linear scaling of $\log{\mtwoh}$ with $\log{\msps}$ \\
$\sigma_{\mathrm{h}}$ & $\mathcal{N}(0.31, 0.04)$ & $0.32\pm0.03$ & Intrinsic scatter in $\log{\mtwoh}$ \\
$\mu_{\sigma,0}$ & $U(1,3)$ & $2.352\pm0.014$ & Mean $\sap$ at $\log{\msps}=11.3$ and average size and halo mass \\
$\beta_{\sigma}$ & $U(0,1)$ & $0.33\pm0.03$ & Dependence of the mean $\log{\sigmaap}$ on $\log{\msps}$ \\
$\xi_{\sigma}$ & $U(-2,0)$ & $-0.45\pm0.09$ & Dependence of the mean $\log{\sigmaap}$ on excess size \\
$\nu_{\sigma}$ & $U(-1,1)$ & $0.06\pm0.05$ & Dependence of the mean $\log{\sigmaap}$ on excess halo mass \\
$\sigma_{\sigma}$ & $U(0,1)$ & $0.035\pm0.009$ & Scatter in $\gamma$ around the mean \\
$\mu_{z_{\mathrm{s}}}$ & $0.48$ (Fixed) & -- & Mean of the effective source redshift distribution \\
$\sigma_{z_{\mathrm{s}}}$ & $0.215$ (Fixed) & -- & Dispersion of the effective source redshift distribution \\
$\theta_0$ & $U(0,3)$ & $0.84 \pm 0.06$ & Lens finding probability parameter \\
$\log{a}$ & $U(-1,3)$ & $1.09 \pm 0.13$ & Lens finding probability parameter
\end{tabular}
\tablefoot{
Model hyper-parameters. Prior choices and inferred values (median and $68\%$ credible region). 
}
\end{table*}

\subsection{Priors}

We assumed a uniform prior on $\log{\asps}$, over the range $(0.0,0.3)$. The lower bound corresponds to a Chabrier IMF, while the upper bound is slightly above the value corresponding to a Salpeter IMF, which is $\log{\asps}=0.25$.
For the contraction parameter $\epsilon$, we assumed a uniform prior over the range $(0,1)$. The lower bound corresponds to NFW dark matter halos, while the upper bound describes maximally contracted halos.
We also assumed uniform priors on the parameters describing the fundamental hyper-plane of \Eref{eq:sigmadist}.
For the halo mass parameters we assumed Gaussian priors, as explained in section \ref{ssec:popmodel}.
We assumed a flat prior on the lens finding parameters $\theta_0$ and $\log{a}$.
Finally, we fixed the source redshift distribution parameters to $\mu_{\zsource} = 0.48$ and $\sigma_{\zsource} = 0.215$, which are the values measured in Paper I. This allowed us to greatly simplify the calculation of the integrals of \Eref{eq:pslnorm}. In Paper I we found that none of the galaxy mass parameters correlated with them, hence we believe that this choice does not introduce any significant bias.
These prior choices are summarised in \Tref{tab:results}.

\section{Results}\label{sect:results}

\subsection{Mass parameters}

\begin{figure}
\includegraphics[width=\columnwidth]{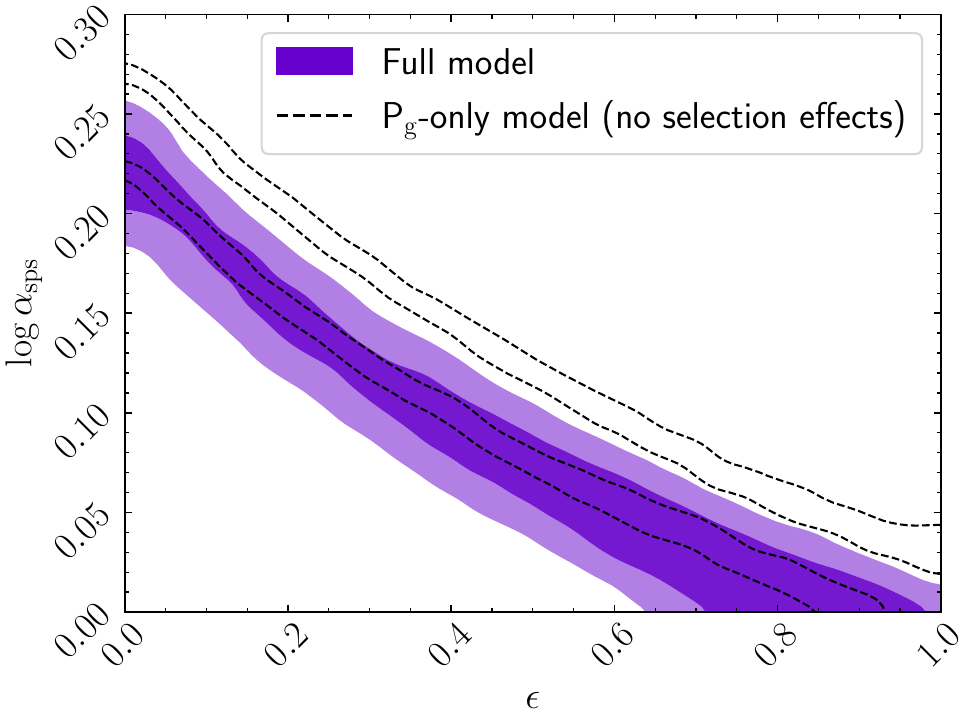}
\caption{
Posterior probability in the dark matter contraction and stellar population synthesis parameters.
The fiducial inference is shown by the solid think contours.
Filled contours show the inference obtained by ignoring selection effects altogether.
Contour levels correspond to $68\%$ and $95\%$ enclosed probability.
\label{fig:epsasps}
}
\end{figure}

\Fref{fig:epsasps} shows the posterior probability in the two main parameters of the model: $\asps$ and $\epsilon$.
Individually, the two are essentially unconstrained, as they are degenerate with each other.
This was expected: increasing the efficiency of dark matter contraction raises the inner dark matter density, which, in order to fit the Einstein radius of a lens, needs to be compensated by a corresponding decrease in stellar mass.
In the limit $\epsilon=0.8$, the data requires $\log{\asps}\approx0$, which is the value of $\asps$ corresponding to a Chabrier IMF.
At the opposite limit, models with $\epsilon=0$ and $0.20 \lesssim \log{\asps} \lesssim0.25$ are allowed by the data.

In order to gauge the importance of selection effects, we also fitted a model with no lensing selection-related terms, essentially dropping $\prsourceeff$ and $\psel$ from \Eref{eq:one}. The resulting inference on $(\epsilon,\asps)$ is shown in \Fref{fig:epsasps} as dashed contours.
At fixed $\epsilon$, $\asps$ is overestimated by approximately $0.04$~dex, an amount that is larger than the $1\sigma$ uncertainty. This shows that selection effects are an important component of strong lensing measurements, already at the current level of precision.

\subsection{Velocity dispersion}

\begin{figure*}
\sidecaption
 \includegraphics[width=12cm]{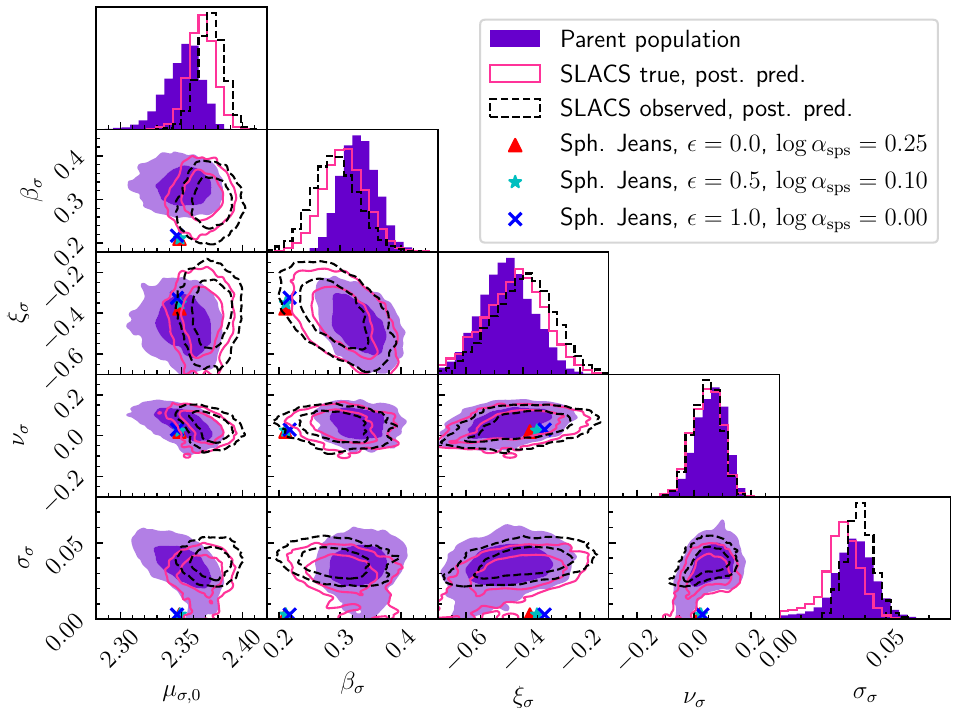}
 \caption{Posterior probability of the parameters describing the distribution in velocity dispersion. These are defined in \Eref{eq:sigmadist} and \Eref{eq:sigmamu}. 
Filled purple contours: posterior probability of the model. 
Solid pink contours: posterior predicted fundamental hyperplane of the SLACS lenses.
Dashed black contours: posterior predicted fundamental hyperplane of the SLACS lenses, based on the observed (noisy) velocity dispersion.
Contour levels indicate regions of $68\%$ and $95\%$ enclosed probabilities.
The three points are values of the parameters obtained by means of dynamical modelling via the isotropic spherical Jeans equation, for different pairs of values of $(\asps,\epsilon)$.
\label{fig:sigmapars}
}
\end{figure*}

\Fref{fig:sigmapars} shows the posterior probability in the parameters describing the distribution in velocity dispersion, or the fundamental hyper-plane.
In order to interpret these results, we also predicted the values of these parameters by using the spherical Jeans equation, assuming orbital isotropy, for a few pairs of values of $(\epsilon,\asps)$ that are consistent with the data.
The Jeans model prediction can match the inferred values of $\mu_{\sigma,0}$, $\xi_{\sigma}$ and $\nu_\sigma$, which describe, respectively, the average velocity dispersion of galaxies at $\log{\msps}=11.3$, and the scaling of velocity dispersion with size and halo mass. However, the Jeans model underpredicts the value of $\beta_\sigma$: the data favour a significantly steeper scaling of velocity dispersion with stellar mass.
This mismatch could be an indication of a mass-dependent $\asps$, which would be consistent with previous findings from lensing and dynamics analyses \citep{Tre++10,Son++15,Pos++15}. But a trend of anisotropy with stellar mass could also explain the discrepancy.

Spherical and isotropic Jeans models predict $\sigma_\sigma=0$ by construction, as the choice of mass profile determines the velocity dispersion uniquely.
A fundamental hyper-plane with zero intrinsic scatter is not completely ruled out by the SLACS data, but is highly disfavoured. We do expect some non-zero scatter in velocity dispersion, for the reason that galaxies are not spherical, their orientation is random, and their anisotropy profile is unlikely to be the same across the whole population.

As we explained in \Sref{sect:intro}, one of the goals of this work is to understand whether the velocity dispersion of the SLACS lenses is biased with respect to their parent population, and whether the observed values of $\sigmaap$ are biased with respect to the truth.
We did so by first obtaining the fundamental hyper-plane parameters of SLACS-like samples of lenses from posterior prediction. We drew sets of values of the model parameters from the posterior probability; then, at each draw, we generated sets of SLACS-like lenses and fitted the distribution in velocity dispersion of \Eref{eq:sigmadist} to these mock data. The resulting distribution in fundamental hyper-plane parameters is shown in \Fref{fig:sigmapars} as pink solid contours.
The posterior predicted fundamental hyper-plane of the SLACS lenses has a larger value of $\mu_{\sigma,0}$ than that of the parent population, meaning a larger velocity dispersion at fixed stellar mass, size and halo mass. 
This bias was not present in the model of Paper I, because the mass distribution determined uniquely the velocity dispersion, by means of the stellar dynamics model. In other words, in the context of Paper I, SLACS lenses differed from parent population galaxies in terms of their inner structure, but had the same velocity dispersion at fixed density profile. Here we allowed for scatter between the mass properties of the lens and the velocity dispersion, therefore traded part of the bias on the mass into a bias on the velocity dispersion.
We believe this new model to be more realistic, as it does not rely on the assumptions of spherical symmetry and orbital isotropy.

In Paper I we found that the velocity dispersion of the SLACS lenses is systematically overestimated.
To verify whether that finding holds with this model as well, we fitted the model of \Eref{eq:sigmadist} directly to the posterior predicted noisy $\sapobs$, instead of the noiseless $\sap$.
The resulting distribution is shown in \Fref{fig:sigmapars} as black dashed contours. As expected from the results of Paper I, the value of $\mu_{\sigma,0}$ obtained in this way is yet larger than that based on the true distribution, meaning that the observed velocity dispersion of SLACS lenses is on average larger than the true value. 

\begin{figure}
\includegraphics[width=\columnwidth]{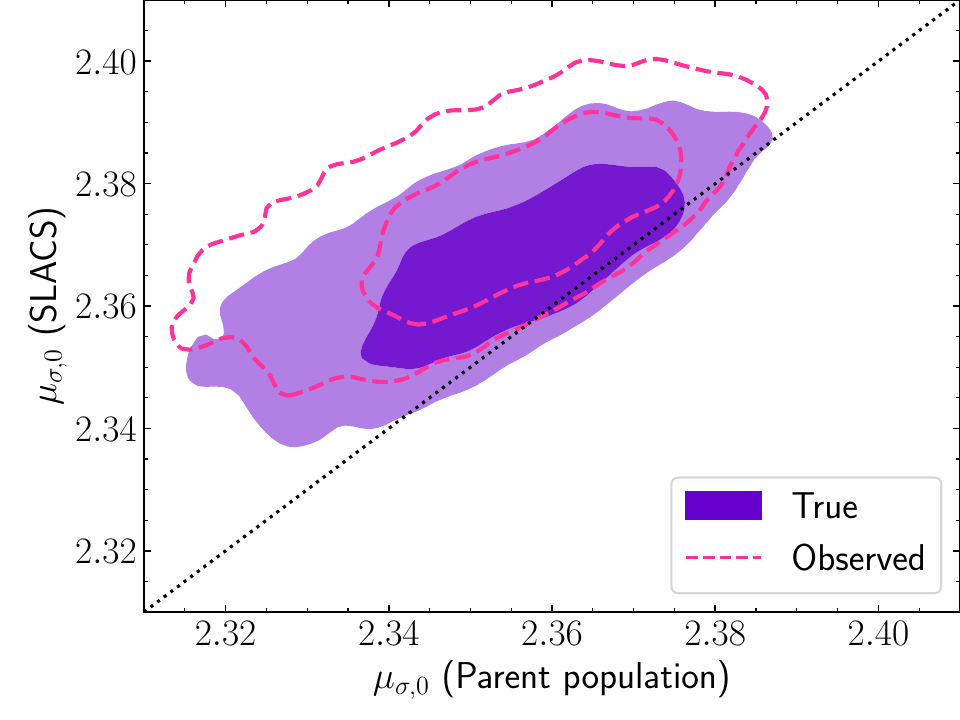}
\caption{
Posterior predicted distribution in the observed and true $\mu_{\sigma,0}$ of SLACS lenses, as a function of the corresponding value of the parent population. The quantity $\mu_{\sigma,0}$ describes the average $\log{\sigmaap}$ of galaxies with $\log{\msps}=11.3$, average size and average halo mass for their stellar mass.
\label{fig:musigmapost}
}
\end{figure}

In order to better illustrate the difference between these various values of $\mu_{\sigma,0}$ and account for their covariance, we show in \Fref{fig:musigmapost} the distribution of $\mu_{\sigma,0}$ of the true and observed velocity dispersion of the SLACS lenses, as a function of the corresponding value for the parent population.
Averaged over the uncertainties, our model predicts that SLACS lenses have a $3\%$ larger velocity dispersion, compared to that of galaxies of the same stellar mass, size and halo mass, and their observed velocity dispersion is a further $2\%$ higher.
Hence, intrinsic and observational bias contribute roughly equally to the bias in velocity dispersion of the SLACS lenses.
%



The Markov Chain Monte Carlo of the two inferences are available online\footnote{\url{https://github.com/astrosonnen/strong_lensing_tools/tree/main/papers/slacs_debiased_2}}, along with posterior predicted mock samples of the galaxy and strong lens population. These mocks can be used to verify, further investigate and better quantify any difference between SLACS and the parent population.

\subsection{Goodness of fit}

\begin{figure*}
\sidecaption
\includegraphics[width=12cm]{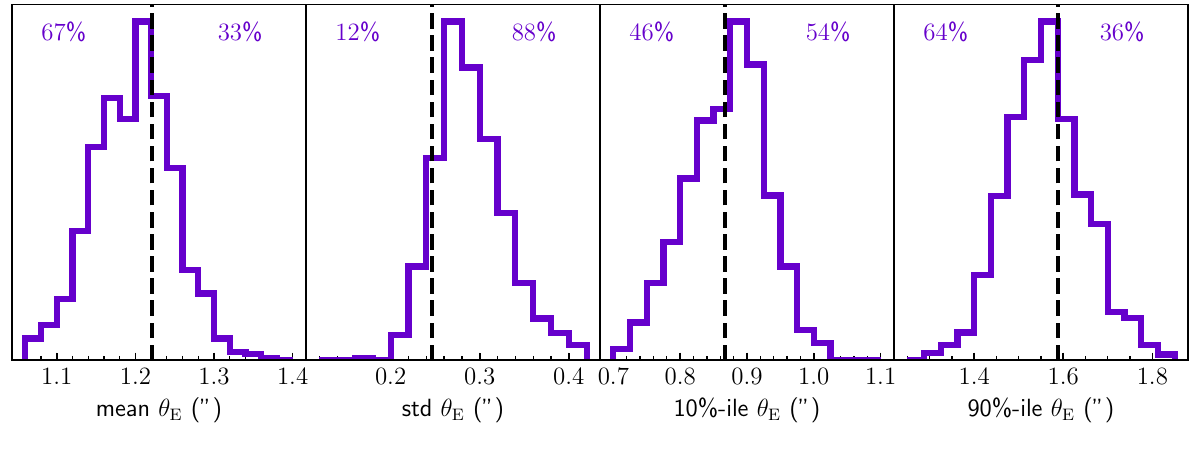}
\caption{
Posterior predictive test of goodness-of-fit.
Vertical lines indicate the observed values of the four test quantities.
In each panel, the percentage to the left and right of the vertical line indicate the fraction of posterior predicted samples for which the mock test quantity is smaller or larger than observed.
\label{fig:pptest}
}
\end{figure*}

We checked the goodness of fit of the model by running posterior predictive tests. This means choosing a series of scalar test quantities derived from the data that we wish the model to reproduce, generating mock observations by sampling from the posterior probability, and measuring the probability of the model to produce more extreme values than the test quantities. A very low or very large probability indicates that the model is unlikely to reproduce those aspects of the data.
We defined the test quantities by focusing on the Einstein radius. We chose the average, the standard deviation, the $10\%$- and the $90\%$-ile of the observed distribution in Einstein radius.
\Fref{fig:pptest} shows the posterior predicted distribution of the four test quantities.
In all cases, the observed values are well within the posterior predicted distributions, hence the model is able to reproduce these data.
%

\subsection{Prediction of other lensing observables}

Since the SLACS sample and the data used for our analysis are not sufficient to individually determine $\asps$ and $\epsilon$, it can be useful to investigate whether current or future data from SLACS or SLACS-like samples can break this degeneracy. 
We did this, too, with posterior prediction. 
We first focused on the lensing-only power-law slope, which is the value of the density slope $\gammapl$ that is obtained by fitting a lens model with a power-law radial density profile, $\rho(r)\propto r^{-\gammapl}$, to imaging data of a strong lens.
\citet{Sha++21}, \citet{Eth++22} and \citet{Tan++24} measured $\gammapl$ for $21$, $29$, and $28$ of the SLACS lenses included in this work, respectively.
Some of these measurements could, in principle, be used as additional constraints. But since our model is not a power-law, we must first understand what exact property of the lenses do the $\gammapl$ capture.
\citet{Son18} and \citet{Koc20} argued that, when a power-law model is fitted to well-resolved, high signal-to-noise ratio lensed images around nearly circularly symmetric lenses, $\gammapl$ is constrained by the ratio of radial magnification between the main arc and its counter-image.
\citet{Gom++23} showed by means of simulations that this is the case, provided that the fitted model has sufficient flexibility in the azimuthal direction.
The radial magnification ratio of images produced by a circular lens can be expressed as a function of the following ratio of radial derivatives of the lensing potential $\psi$, evaluated at the Einstein radius:
\begin{equation}\label{eq:xirad}
\xi_{\mathrm{rad}} = \tein\frac{\psi_{\mathrm{E}}'''}{1 - \psi_{\mathrm{E}}''}.
\end{equation} 
For a power-law model, 
\begin{equation}\label{eq:xipl}
\xi_{\mathrm{rad}} = \gammapl - 2.
\end{equation}
Therefore, we can define an equivalent lensing-only power-law slope for our model, by computing $\psi_{\mathrm{E}}''$ and $\psi_{\mathrm{E}}'''$, obtaining $\xi_{\mathrm{rad}}$ from \Eref{eq:xirad}, and setting $\gammapl = \xi_{\mathrm{rad}}+ 2$.

\begin{figure}
\includegraphics[width=\columnwidth]{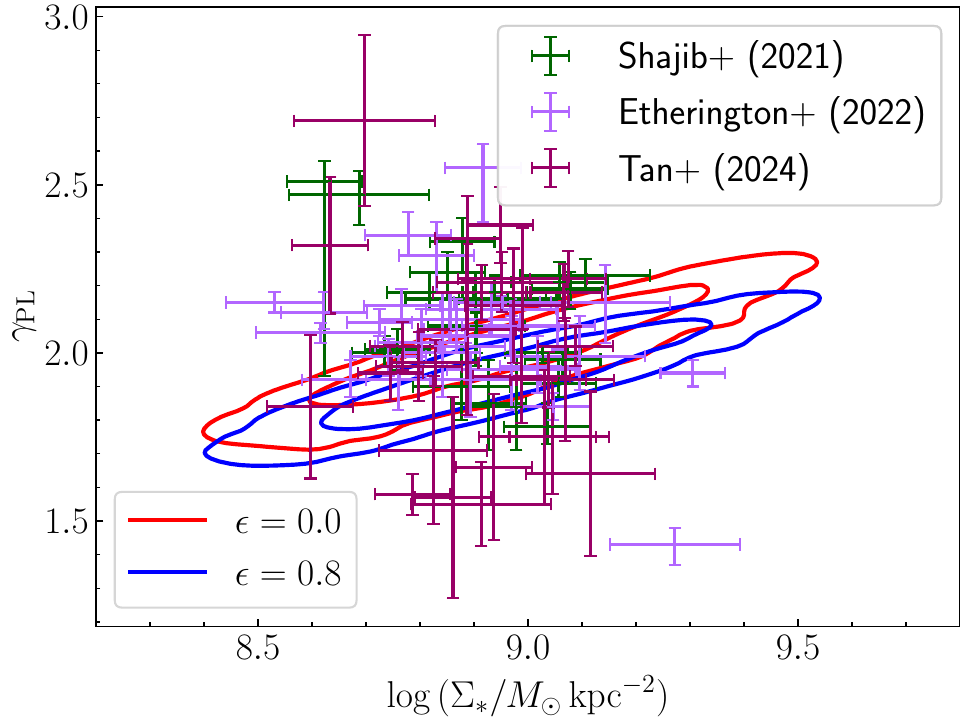}
\caption{
Posterior predicted distribution in the lensing-only power-law slope, as a function of stellar mass density. 
Simulated values of $\gammapl$ have been obtained from \Eref{eq:xipl}.
The two sets of contours correspond to the posteriors obtained by fitting models with a fixed value of $\epsilon$, as indicated in the legend.
Error bars are measurements from \citet{Sha++21}, \citet{Eth++22}, \citet{Tan++24}.
\label{fig:gammapl}
}
\end{figure}

We did this for our model, while fixing the value of the dark matter contraction parameters to two extreme values: $\epsilon=0.0$ and $\epsilon=0.8$. \Fref{fig:gammapl} shows the posterior predicted distribution in $\gammapl$, as a function of the stellar surface mass density, $\sstar=\msps/(2\pi\reff^2)$. The two quantities are positively correlated: this is expected, since $\sstar$ correlates with the total density slope $\gamma$ obtained from lensing and dynamics \citep{Son++13b}, which should be closely related to $\gammapl$.
The $\epsilon=0.0$ and $\epsilon=0.8$ model lie on different regions of the $\sstar-\gammapl$ parameter space. Although there is overlap between the two posterior predicted distributions, most their widths is due to intrinsic scatter within the population of lenses, rather than on the parameter uncertainty: therefore, the two models stand apart when comparing the average $\gammapl$ of the lens population. 
This means that $\gammapl$ observations could already help distinguish between uncontracted or significantly contracted dark matter profiles.
Unfortunately, as \Fref{fig:gammapl} shows, the distribution of current measurements of $\gammapl$ of the SLACS lenses bears very little resemblance to that predicted with our model.
The observed $\gammapl$ appear to be anti-correlated with $\sstar$. This is at odds with lensing and dynamics constrains, as already pointed out by \citet{Eth++23}.

\begin{figure*}
\sidecaption
\includegraphics[width=12cm]{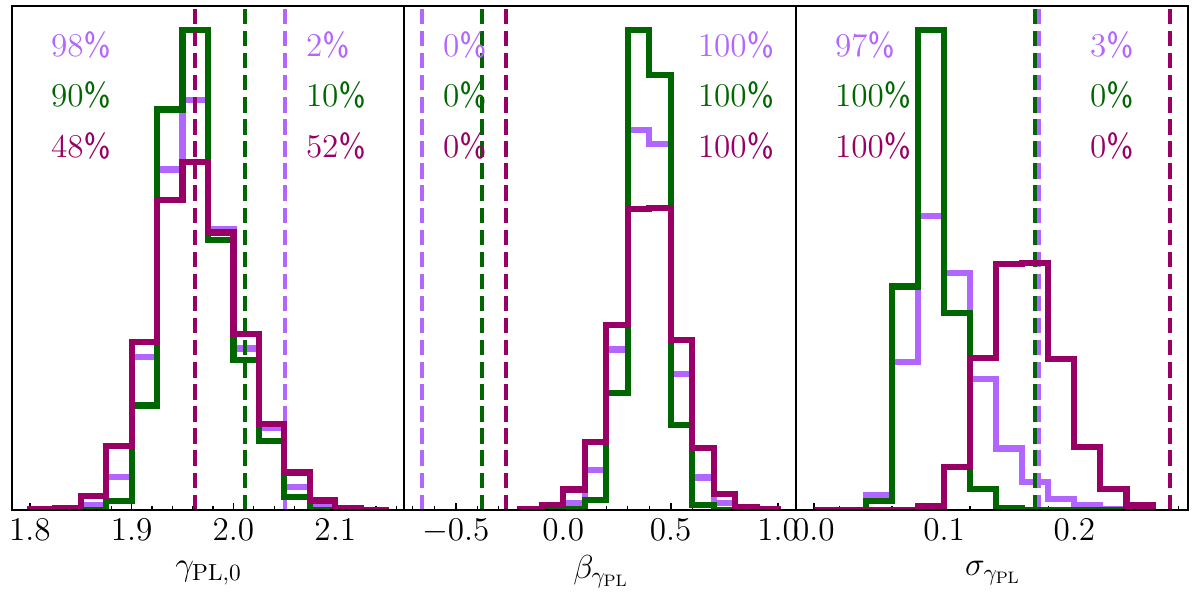}
\caption{
Posterior predicted distribution of the coefficients describing the $\gammapl-\sstar$ relation of SLACS lenses.
The coefficients are defined in \Eref{eq:gammapllin}.
For each coefficient, three distributions are shown, each to be compared with the value obtained by fitting the measurements of \citet{Sha++21}, \citet{Eth++22} and \citet{Tan++24}. The colour of each distribution corresponds to that of the data points in \Fref{fig:gammapl}.
The distributions are different because the observed samples and the observational uncertainties vary from one work to the other.
In each panel, the percentages to the left and right of the vertical lines indicate the fraction of posterior predicted samples for which the mock test quantity is smaller or larger than observed.
\label{fig:gammapltest}
}
\end{figure*}

We quantified the discrepancy with another posterior predictive test, in which we generated noisy measurements of $\gammapl$, fitted them a linear relation of the kind
\begin{equation}\label{eq:gammapllin}
\gammaplobs(\Sigma_*^{(\mathrm{obs})}) = \mathcal{N}_{\mathrm{\gammaplobs}}(\gamma_{\mathrm{PL},0} + \beta_{\gammapl}(\Sigma_*^{(\mathrm{obs})} - 9),\sigma_{\gammapl}^2),
\end{equation}
and counted the fraction of times in which the parameters have more extreme values than those obtained by fitting \Eref{eq:gammapllin} to the observed samples of \citet{Sha++21}, \citet{Eth++22} and \citet{Tan++24}.
The results are shown in \Fref{fig:gammapltest}. Only the intercept $\gamma_{\mathrm{PL},0}$ can be reproduced by the model, and only for two of the datasets. All three of the observed datasets show an anti-correlation between $\gammapl$ and $\sstar$, corresponding to a negative value of $\beta_{\gammapl}$. Out of 10000 posterior predicted mocks, only a handful have a value of $\beta_{\gammapl}$ that is smaller than the observed ones.

This discrepancy could be explained by the model being inaccurate.
However, given the fact that many of the measurements of $\gammapl$ from the different works disagree on individual lenses \citep[see Figure 6 of][]{Tan++24}, we believe that there are problems with the observations, or their interpretation.
One possible source of error could lie in the choice of the model for the measurement of $\gammapl$. \citet{Sha++21}, \citet{Eth++22} and \citet{Tan++24} fitted lens models with elliptical symmetry to the SLACS lenses.
If the true azimuthal structure of the lenses deviates from elliptical symmetry, then the value of $\gammapl$ obtained can no longer interpreted as a measurement of the radial profile via \Eref{eq:xipl}. As argued by \citet{Koc21} and shown in simulations by \citet{vdV++22} and \citet{Gom++23}, fits of elliptical models to lenses with more complex azimuthal structure lead to biases. Other sources of error, which could explain the differences between the three measurements on the same lenses, are the modelling of the point spread function and of the background source \citep[see][ for a discussion]{Gal++24}.
We conclude that, at the present, we cannot rely on existing measurements of $\gammapl$, until it is shown that they are reproducible and robust to the effects mentioned above.
Nevertheless, measurements of the radial magnification ratio on large sample of lenses have the potential to break the degeneracy between the stellar IMF and the dark matter profile, as also shown by \citet{S+C21}.

\begin{figure}
\includegraphics[width=\columnwidth]{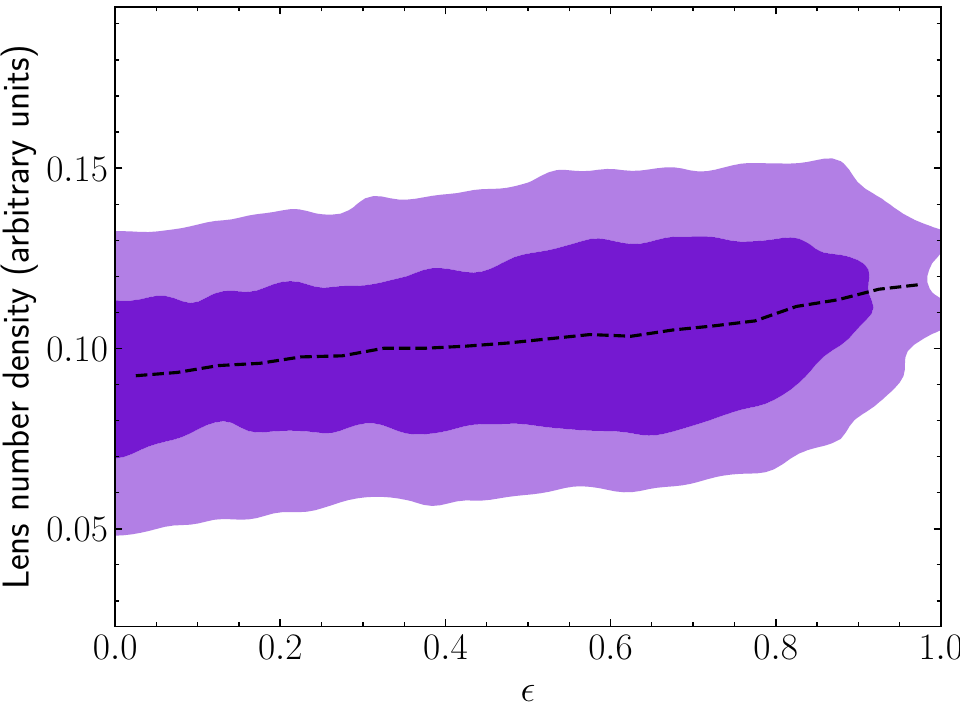}
\caption{
Posterior predicted number density of SLACS lenses, as a function of the contraction efficiency parameter. The dashed line shows the average $\nlens$, obtained by marginalising over the uncertainty on the model parameters.
\label{fig:pslnorm}
}
\end{figure}

Another piece of information that could help break the degeneracy between $\asps$ and $\epsilon$ is the number density of lenses. As shown by \citet{ZSH24}, the number of lenses that are found in a given area of the sky can provide complementary constraints on the structural parameters of the galaxy population.
In the context of our model, the number density of lenses in a given survey is given by the following product:
\begin{equation}
\nlens = \nfgd \nbkg \int d\psilens d\zsource \prlens(\psilens) \prsourceeff(\zsource) g(\psilens,\zsource) \pfind(\psilens,\zsource),
\end{equation}
where $\nfgd$ and $\nbkg$ are the projected number density of foreground galaxies and background sources, respectively, and $g$ is the geometrical component of the lensing cross-section.
We do not have precise estimates of $\nfgd$, which depends on the effective area of the SLACS survey, and especially of $\nbkg$.
Nevertheless, we can study how $\nlens$ varies as a function of the lens population parameters for arbitrary values of $\nfgd$ and $\nbkg$. We computed this quantity by Monte Carlo integration, in the same way as we obtained the normalisation of $\prsleff$.
\Fref{fig:pslnorm} shows the posterior predicted $\nlens$ as a function of $\epsilon$.
The two quantities are positively correlated: when averaged over the uncertainty on the other parameters, the model with $\epsilon=0.8$ predicts a $20\%$ larger $\nlens$ compared to the $\epsilon=0.0$ case.
Although at fixed $\epsilon$ the predicted distribution in $\nlens$ has a relatively large spread, most of this spread is due to the uncertainty on the selection function parameters, in particular $\theta_0$.
This means that, if the selection function was known exactly and both $\nfgd$ and $\nbkg$ were known precisely, a sample of $\sim100$ lenses would allow us to distinguish between the two extreme scenarios allowed by the current constraints. Although this might be difficult to achieve for the SLACS sample, future surveys with better-defined selection criteria might be more fortunate in this regard.

\section{Discussion and conclusions}\label{sect:discuss}

We used the SLACS sample to revisit the issue of the relative contribution of baryonic and dark matter to the mass budget of early-type galaxies, in view of new insight on its selection function obtained in Paper I.
We tried to constrain the stellar mass-to-light ratio, described by the average stellar population synthesis mismatch parameter $\asps$, and the amount of dark matter contraction with respect to an NFW profile, described by the contraction efficiency parameter $\epsilon$. 
Although we were not able to constrain either of them individually, we put a tight bound on their combination that allowed us to rule out large regions in the $\asps-\epsilon$ parameter space.
A stellar IMF heavier than Salpeter (i.e. that produces a larger mass-to-light ratio), corresponding to $\log{\asps} > 0.25$, is disfavoured by the data.
Although the shape of the $\epsilon-\asps$ correlation suggests that larger values of $\asps$ might be allowed if we let $\epsilon$ to extend below zero, which would correspond to expansion of the halo, we have a strong prior against such a solution, because hydrodynamical simulations predict contraction.

At fixed $\epsilon$, our inferred value of $\asps$ is generally lower than that of most other works in the literature. 
For instance, \citet{Sha++21} measured $\log{\asps}=0.26$ on average, for a halo profile with no contraction or expansion, in their analysis of $21$ SLACS lenses. The difference with respect to our measurement, $0.04$~dex, is identical to the shift in the posterior probability due to the inclusion of the terms related to the selection function, hence it can be attributed entirely to it.

We also investigated how the velocity dispersion of the SLACS lenses compares to that of parent population galaxies, at fixed mass profile. We found that SLACS lenses have a $3\%$ larger velocity dispersion, and that their observed velocity dispersion is biased upwards by another $2\%$. 
These biases pose challenges for the use of SLACS lenses in the aid of time-delay cosmography measurements. They imply that, in order to interpret stellar kinematics measurements with a dynamical model, priors that take the SLACS selection function into account must be used.
One might hope to remove at least the observational bias with higher signal-to-noise spectroscopic measurements that supersede the SDSS-based ones on which our analysis is based. However, whether this can be done or not depends on what exact process generated the observational scatter in the SDSS velocity dispersion measurements. In the ideal case of uncorrelated statistical noise, then the uncertainty should go to zero in the limit of high signal-to-noise. However, if systematic effects are at play, such as a mismatch between the stellar templates used for the velocity dispersion fit and the intrinsic spectra of the lenses, correcting for this bias might be difficult, as it might still be present in better quality data.

\citet{Kna++25} have recently obtained spatially-resolved spectroscopic data on $13$ SLACS lenses with the Keck Cosmic Web Imager (KCWI), and compared SDSS-based measurements of $\sigmaap$ with the same quantity obtained from these higher quality data. They found that the SDSS $\sigmaapobs$ are under-estimated by a few percent. This bias goes in the opposite direction than the one predicted by our model. Assuming that the same bias applies to all SDSS spectra of massive early-type galaxies, this does not change the conclusions of our study, as it would simply shift the overall value of $\mu_{\sigma,0}$, but not the relative difference between the SLACS lenses and the parent population. 

Finally, we explored ways to break the degeneracy between $\epsilon$ and $\asps$ with future lensing data. One observable quantity that could help in this regard is the lensing-only density slope, $\gammapl$, to be obtained by measuring the ratio of radial magnification between the multiple images. Although such measurements already exist for a fraction of the SLACS lenses, they are completely inconsistent with our model. In particular, while the model predicts a positive correlation between stellar mass density and $\gammapl$, the observations show an anti-correlation.
While we have reasons to believe that there are problems with some of these measurements, it would be interesting to also explore whether it is possible to find models that better match the observations. We leave that to future work.

In summary, this analysis constitutes a shift in strategy in the analysis of samples of strong lenses, moving from joint lensing and dynamics to purely lensing-based modelling. This choice allowed us to focus exclusively on projected quantities, avoiding the need to predict difficult properties to measure, such as the stellar anisotropy and the three-dimensional structure of lens galaxies.
We modelled the distribution of stellar and dark matter of the SLACS lenses, together with their selection function.
This gave us a twofold advantage: it allowed us to correct for selection effects, which we found to have a significant impact when compared to the observational uncertainty, and to incorporate weak lensing information from the parent population of SLACS lens galaxies.
The approach used in this work can serve as a solid foundation on which to build our understanding of galaxy structure with upcoming lensing data. Combined with a larger number of lenses, a yet better understanding of selection effects, and number density or reliable radial magnification information, it will make possible to break the degeneracy between the stellar IMF and the dark matter density profile in early-type galaxies.

\bibliographystyle{aa}
\bibliography{references}

\begin{thebibliography}{37}
\expandafter\ifx\csname natexlab\endcsname\relax\def\natexlab#1{#1}\fi

\bibitem[{{Aihara} {et~al.}(2018{\natexlab{a}}){Aihara}, {Arimoto},
  {Armstrong}, {Arnouts}, {Bahcall}, {Bickerton}, {Bosch}, {Bundy}, {Capak},
  {Chan}, {Chiba}, {Coupon}, {Egami}, {Enoki}, {Finet}, {Fujimori}, {Fujimoto},
  {Furusawa}, {Furusawa}, {Goto}, {Goulding}, {Greco}, {Greene}, {Gunn},
  {Hamana}, {Harikane}, {Hashimoto}, {Hattori}, {Hayashi}, {Hayashi},
  {He{\l}miniak}, {Higuchi}, {Hikage}, {Ho}, {Hsieh}, {Huang}, {Huang},
  {Ikeda}, {Imanishi}, {Inoue}, {Iwasawa}, {Iwata}, {Jaelani}, {Jian},
  {Kamata}, {Karoji}, {Kashikawa}, {Katayama}, {Kawanomoto}, {Kayo}, {Koda},
  {Koike}, {Kojima}, {Komiyama}, {Konno}, {Koshida}, {Koyama}, {Kusakabe},
  {Leauthaud}, {Lee}, {Lin}, {Lin}, {Lupton}, {Mandelbaum}, {Matsuoka},
  {Medezinski}, {Mineo}, {Miyama}, {Miyatake}, {Miyazaki}, {Momose}, {More},
  {More}, {Moritani}, {Moriya}, {Morokuma}, {Mukae}, {Murata}, {Murayama},
  {Nagao}, {Nakata}, {Niida}, {Niikura}, {Nishizawa}, {Obuchi}, {Oguri},
  {Oishi}, {Okabe}, {Okamoto}, {Okura}, {Ono}, {Onodera}, {Onoue}, {Osato},
  {Ouchi}, {Price}, {Pyo}, {Sako}, {Sawicki}, {Shibuya}, {Shimasaku},
  {Shimono}, {Shirasaki}, {Silverman}, {Simet}, {Speagle}, {Spergel},
  {Strauss}, {Sugahara}, {Sugiyama}, {Suto}, {Suyu}, {Suzuki}, {Tait},
  {Takada}, {Takata}, {Tamura}, {Tanaka}, {Tanaka}, {Tanaka}, {Tanaka},
  {Terai}, {Terashima}, {Toba}, {Tominaga}, {Toshikawa}, {Turner}, {Uchida},
  {Uchiyama}, {Umetsu}, {Uraguchi}, {Urata}, {Usuda}, {Utsumi}, {Wang}, {Wang},
  {Wong}, {Yabe}, {Yamada}, {Yamanoi}, {Yasuda}, {Yeh}, {Yonehara}, \&
  {Yuma}}]{Aih++18a}
{Aihara}, H., {Arimoto}, N., {Armstrong}, R., {et~al.} 2018{\natexlab{a}},
  \pasj, 70, S4

\bibitem[{{Aihara} {et~al.}(2018{\natexlab{b}}){Aihara}, {Armstrong},
  {Bickerton}, {Bosch}, {Coupon}, {Furusawa}, {Hayashi}, {Ikeda}, {Kamata},
  {Karoji}, {Kawanomoto}, {Koike}, {Komiyama}, {Lang}, {Lupton}, {Mineo},
  {Miyatake}, {Miyazaki}, {Morokuma}, {Obuchi}, {Oishi}, {Okura}, {Price},
  {Takata}, {Tanaka}, {Tanaka}, {Tanaka}, {Uchida}, {Uraguchi}, {Utsumi},
  {Wang}, {Yamada}, {Yamanoi}, {Yasuda}, {Arimoto}, {Chiba}, {Finet},
  {Fujimori}, {Fujimoto}, {Furusawa}, {Goto}, {Goulding}, {Gunn}, {Harikane},
  {Hattori}, {Hayashi}, {He{\l}miniak}, {Higuchi}, {Hikage}, {Ho}, {Hsieh},
  {Huang}, {Huang}, {Imanishi}, {Iwata}, {Jaelani}, {Jian}, {Kashikawa},
  {Katayama}, {Kojima}, {Konno}, {Koshida}, {Kusakabe}, {Leauthaud}, {Lee},
  {Lin}, {Lin}, {Mandelbaum}, {Matsuoka}, {Medezinski}, {Miyama}, {Momose},
  {More}, {More}, {Mukae}, {Murata}, {Murayama}, {Nagao}, {Nakata}, {Niida},
  {Niikura}, {Nishizawa}, {Oguri}, {Okabe}, {Ono}, {Onodera}, {Onoue}, {Ouchi},
  {Pyo}, {Shibuya}, {Shimasaku}, {Simet}, {Speagle}, {Spergel}, {Strauss},
  {Sugahara}, {Sugiyama}, {Suto}, {Suzuki}, {Tait}, {Takada}, {Terai}, {Toba},
  {Turner}, {Uchiyama}, {Umetsu}, {Urata}, {Usuda}, {Yeh}, \&
  {Yuma}}]{Aih++18b}
{Aihara}, H., {Armstrong}, R., {Bickerton}, S., {et~al.} 2018{\natexlab{b}},
  \pasj, 70, S8

\bibitem[{{Auger} {et~al.}(2009){Auger}, {Treu}, {Bolton}, {Gavazzi},
  {Koopmans}, {Marshall}, {Bundy}, \& {Moustakas}}]{Aug++09}
{Auger}, M.~W., {Treu}, T., {Bolton}, A.~S., {et~al.} 2009, \apj, 705, 1099

\bibitem[{{Auger} {et~al.}(2010{\natexlab{a}}){Auger}, {Treu}, {Bolton},
  {Gavazzi}, {Koopmans}, {Marshall}, {Moustakas}, \& {Burles}}]{Aug++10a}
{Auger}, M.~W., {Treu}, T., {Bolton}, A.~S., {et~al.} 2010{\natexlab{a}}, \apj,
  724, 511

\bibitem[{{Auger} {et~al.}(2010{\natexlab{b}}){Auger}, {Treu}, {Gavazzi},
  {Bolton}, {Koopmans}, \& {Marshall}}]{Aug++10b}
{Auger}, M.~W., {Treu}, T., {Gavazzi}, R., {et~al.} 2010{\natexlab{b}}, \apjl,
  721, L163

\bibitem[{{Barnab{\`e}} {et~al.}(2011){Barnab{\`e}}, {Czoske}, {Koopmans},
  {Treu}, \& {Bolton}}]{Bar++11}
{Barnab{\`e}}, M., {Czoske}, O., {Koopmans}, L. V.~E., {Treu}, T., \& {Bolton},
  A.~S. 2011, \mnras, 415, 2215

\bibitem[{{Birrer} {et~al.}(2020){Birrer}, {Shajib}, {Galan}, {Millon}, {Treu},
  {Agnello}, {Auger}, {Chen}, {Christensen}, {Collett}, {Courbin}, {Fassnacht},
  {Koopmans}, {Marshall}, {Park}, {Rusu}, {Sluse}, {Spiniello}, {Suyu},
  {Wagner-Carena}, {Wong}, {Barnab{\`e}}, {Bolton}, {Czoske}, {Ding},
  {Frieman}, \& {Van de Vyvere}}]{Bir++20}
{Birrer}, S., {Shajib}, A.~J., {Galan}, A., {et~al.} 2020, \aap, 643, A165

\bibitem[{{Blumenthal} {et~al.}(1986){Blumenthal}, {Faber}, {Flores}, \&
  {Primack}}]{Blu86}
{Blumenthal}, G.~R., {Faber}, S.~M., {Flores}, R., \& {Primack}, J.~R. 1986,
  \apj, 301, 27

\bibitem[{{Bolton} {et~al.}(2006){Bolton}, {Burles}, {Koopmans}, {Treu}, \&
  {Moustakas}}]{Bol++06}
{Bolton}, A.~S., {Burles}, S., {Koopmans}, L. V.~E., {Treu}, T., \&
  {Moustakas}, L.~A. 2006, \apj, 638, 703

\bibitem[{{Cautun} {et~al.}(2020){Cautun}, {Ben{\'\i}tez-Llambay}, {Deason},
  {Frenk}, {Fattahi}, {G{\'o}mez}, {Grand}, {Oman}, {Navarro}, \&
  {Simpson}}]{Cau++20}
{Cautun}, M., {Ben{\'\i}tez-Llambay}, A., {Deason}, A.~J., {et~al.} 2020,
  \mnras, 494, 4291

\bibitem[{{Duffy} {et~al.}(2010){Duffy}, {Schaye}, {Kay}, {Dalla Vecchia},
  {Battye}, \& {Booth}}]{Duf++10}
{Duffy}, A.~R., {Schaye}, J., {Kay}, S.~T., {et~al.} 2010, \mnras, 405, 2161

\bibitem[{{Dutton} \& {Macci{\`o}}(2014)}]{D+M14}
{Dutton}, A.~A. \& {Macci{\`o}}, A.~V. 2014, \mnras, 441, 3359

\bibitem[{{Dutton} {et~al.}(2007){Dutton}, {van den Bosch}, {Dekel}, \&
  {Courteau}}]{Dut++07}
{Dutton}, A.~A., {van den Bosch}, F.~C., {Dekel}, A., \& {Courteau}, S. 2007,
  \apj, 654, 27

\bibitem[{{Etherington} {et~al.}(2022){Etherington}, {Nightingale}, {Massey},
  {Cao}, {Robertson}, {Amorisco}, {Amvrosiadis}, {Cole}, {Frenk}, {He}, {Li},
  \& {Tam}}]{Eth++22}
{Etherington}, A., {Nightingale}, J.~W., {Massey}, R., {et~al.} 2022, \mnras,
  517, 3275

\bibitem[{{Etherington} {et~al.}(2023){Etherington}, {Nightingale}, {Massey},
  {Robertson}, {Cao}, {Amvrosiadis}, {Cole}, {Frenk}, {He}, {Lagattuta},
  {Lange}, \& {Li}}]{Eth++23}
{Etherington}, A., {Nightingale}, J.~W., {Massey}, R., {et~al.} 2023, \mnras,
  521, 6005

\bibitem[{{Galan} {et~al.}(2024){Galan}, {Vernardos}, {Minor}, {Sluse}, {Van de
  Vyvere}, \& {Gomer}}]{Gal++24}
{Galan}, A., {Vernardos}, G., {Minor}, Q., {et~al.} 2024, \aap, 692, A87

\bibitem[{{Gnedin} {et~al.}(2004){Gnedin}, {Kravtsov}, {Klypin}, \&
  {Nagai}}]{Gne++04}
{Gnedin}, O.~Y., {Kravtsov}, A.~V., {Klypin}, A.~A., \& {Nagai}, D. 2004, \apj,
  616, 16

\bibitem[{{Gomer} {et~al.}(2023){Gomer}, {Sluse}, {Van de Vyvere}, {Birrer},
  {Shajib}, \& {Courbin}}]{Gom++23}
{Gomer}, M.~R., {Sluse}, D., {Van de Vyvere}, L., {et~al.} 2023, \aap, 679,
  A128

\bibitem[{{Hyde} \& {Bernardi}(2009)}]{H+B09}
{Hyde}, J.~B. \& {Bernardi}, M. 2009, \mnras, 394, 1978

\bibitem[{{Knabel} {et~al.}(2024){Knabel}, {Treu}, {Cappellari}, {Shajib},
  {Chen}, \& {Bennert}}]{Kna++25}
{Knabel}, S., {Treu}, T., {Cappellari}, M., {et~al.} 2024, arXiv e-prints,
  arXiv:2409.10631

\bibitem[{{Kochanek}(2020)}]{Koc20}
{Kochanek}, C.~S. 2020, \mnras, 493, 1725

\bibitem[{{Kochanek}(2021)}]{Koc21}
{Kochanek}, C.~S. 2021, \mnras, 501, 5021

\bibitem[{{Koopmans} {et~al.}(2006){Koopmans}, {Treu}, {Bolton}, {Burles}, \&
  {Moustakas}}]{Koo++06}
{Koopmans}, L. V.~E., {Treu}, T., {Bolton}, A.~S., {Burles}, S., \&
  {Moustakas}, L.~A. 2006, \apj, 649, 599

\bibitem[{{Posacki} {et~al.}(2015){Posacki}, {Cappellari}, {Treu},
  {Pellegrini}, \& {Ciotti}}]{Pos++15}
{Posacki}, S., {Cappellari}, M., {Treu}, T., {Pellegrini}, S., \& {Ciotti}, L.
  2015, \mnras, 446, 493

\bibitem[{{Schaller} {et~al.}(2015){Schaller}, {Frenk}, {Bower}, {Theuns},
  {Jenkins}, {Schaye}, {Crain}, {Furlong}, {Dalla Vecchia}, \&
  {McCarthy}}]{Sch++15}
{Schaller}, M., {Frenk}, C.~S., {Bower}, R.~G., {et~al.} 2015, \mnras, 451,
  1247

\bibitem[{{Shajib} {et~al.}(2021){Shajib}, {Treu}, {Birrer}, \&
  {Sonnenfeld}}]{Sha++21}
{Shajib}, A.~J., {Treu}, T., {Birrer}, S., \& {Sonnenfeld}, A. 2021, \mnras,
  503, 2380

\bibitem[{{Sheu} {et~al.}(2024){Sheu}, {Shajib}, {Treu}, {Sonnenfeld},
  {Birrer}, {Cappellari}, {Oldham}, \& {Tan}}]{She++25}
{Sheu}, W., {Shajib}, A.~J., {Treu}, T., {et~al.} 2024, arXiv e-prints,
  arXiv:2408.10316

\bibitem[{{Sonnenfeld}(2018)}]{Son18}
{Sonnenfeld}, A. 2018, \mnras, 474, 4648

\bibitem[{{Sonnenfeld}(2024)}]{Son24}
{Sonnenfeld}, A. 2024, \aap, 690, A325, (Paper I)

\bibitem[{{Sonnenfeld} \& {Cautun}(2021)}]{S+C21}
{Sonnenfeld}, A. \& {Cautun}, M. 2021, \aap, 651, A18

\bibitem[{{Sonnenfeld} {et~al.}(2018){Sonnenfeld}, {Leauthaud}, {Auger},
  {Gavazzi}, {Treu}, {More}, \& {Komiyama}}]{Son++18}
{Sonnenfeld}, A., {Leauthaud}, A., {Auger}, M.~W., {et~al.} 2018, \mnras, 481,
  164

\bibitem[{{Sonnenfeld} {et~al.}(2013){Sonnenfeld}, {Treu}, {Gavazzi}, {Suyu},
  {Marshall}, {Auger}, \& {Nipoti}}]{Son++13b}
{Sonnenfeld}, A., {Treu}, T., {Gavazzi}, R., {et~al.} 2013, \apj, 777, 98

\bibitem[{{Sonnenfeld} {et~al.}(2015){Sonnenfeld}, {Treu}, {Marshall}, {Suyu},
  {Gavazzi}, {Auger}, \& {Nipoti}}]{Son++15}
{Sonnenfeld}, A., {Treu}, T., {Marshall}, P.~J., {et~al.} 2015, \apj, 800, 94

\bibitem[{{Tan} {et~al.}(2024){Tan}, {Shajib}, {Birrer}, {Sonnenfeld}, {Treu},
  {Wells}, {Williams}, {Buckley-Geer}, {Drlica-Wagner}, \& {Frieman}}]{Tan++24}
{Tan}, C.~Y., {Shajib}, A.~J., {Birrer}, S., {et~al.} 2024, \mnras, 530, 1474

\bibitem[{{Treu} {et~al.}(2010){Treu}, {Auger}, {Koopmans}, {Gavazzi},
  {Marshall}, \& {Bolton}}]{Tre++10}
{Treu}, T., {Auger}, M.~W., {Koopmans}, L. V.~E., {et~al.} 2010, \apj, 709,
  1195

\bibitem[{{Van de Vyvere} {et~al.}(2022){Van de Vyvere}, {Sluse}, {Gomer}, \&
  {Mukherjee}}]{vdV++22}
{Van de Vyvere}, L., {Sluse}, D., {Gomer}, M.~R., \& {Mukherjee}, S. 2022,
  \aap, 663, A179

\bibitem[{{Zhou} {et~al.}(2024){Zhou}, {Sonnenfeld}, \& {Hoekstra}}]{ZSH24}
{Zhou}, Q., {Sonnenfeld}, A., \& {Hoekstra}, H. 2024, \aap, 690, A390

\end{thebibliography}

\begin{acknowledgements}
This work was supported by the National Key R\&D Program of China (No. 2023YFA1607800, 2023YFA1607802).

\end{acknowledgements}

\end{document}